\documentclass[12pt]{article}
\begin{document}
\pagestyle{plain}
\title{Time Evolution as a Gauge Transformation: \\
$x^5$-dependent Cosmological Solution in 5d Kaluza-Klein
theory\footnote{This research was supported by the Kyungsung
University research grants in 2000}}
\author{Gyeong Yun Jun, Pyung Seong Kwon
\footnote{E-mail:bskwon@star.kyungsung.ac.kr}\\
{\small Department of Physics, Kyunsung University, Pusan 608-736,
Korea}}
\date{}
\maketitle
\thispagestyle{empty}
\begin{abstract}
We discuss a new feature of the 5d Kaluza-Klein cosmology. For
that purpose, we obtain the simplest $x^5$-dependent solution
which, in the reduced description, is associated with a
radiation-dominated Robertson-Walker universe, and also can be
regarded as an extension of the Schwarzschild solution. This
solution enables us to deduce an important result that an evolving
universe is related with a static universe by the gauge
transformation, i.e., they are gauge equivalent. This means that
having a different universe simply corresponds to choosing a
different gauge.
\end{abstract}
\medskip
\begin{center}
\end{center}
\newpage
\baselineskip 8.5mm

\setcounter{page}{1}
 Birkhoff's theorem states that every spherically symmetric
 solution - regardless of whether it is static or dynamic - to the
 vacuum Einstein equations is essentially the Schwarzschild
 solution\footnote{This is not quite true in the Kaluza-Klein
 theory. See for instance ref.\cite{gro}.}. So, it is not 
surprising that
 the Schwarzschild solution has a cosmological interpretation. The
 5d Schwarzschild  solution has indeed a
 cosmological interpretation describing time evolution of the 4d
 isotropic, homogeneous (Robertson-Walker) universe \cite{beh}.
 One of reasons for considering such an embedding is
because it may provide the possibility of avoiding cosmological
singularities arising in the conventional Robertson-Walker
cosmology. It is known that certain genuine singularities in four
dimensions can be resolved by simply going to higher dimensions
\cite{gib}.

Apart from this, there has recently been proposed a new mechanism
\cite{ark} for solving the hierarchy problem based on the
assumption that the conventional Planck scale $M_{pl} \sim 10^
{19}GeV$ is essentially not the fundamental scale in nature;
$M_{pl} $ is simply an effective constant determined by the
Electroweak scale $M_{EW}$(which is assumed to be the only
short-distance fundamental scale) and the volume (or curvature
\cite{ran}) of the extra dimensions. This assumption then leads to
the result that the hierarchy between $M_{pl}$ and $M_{EW}$ can be
eliminated by taking the extra dimensions to be very large. The
important consequence of this is that Kaluza-Klein excitations are
not insensible anymore, and their effects become important in the
theory. This in turn implies that the dependence of the metric
fields on the extra dimensions is not to be neglected ; rather,
they become crucial \cite{gid}. In this letter, we are examining
the simplest $x^5$- dependent ($x^5$ being the fifth coordinate)
cosmological solution of the standard 5d
 Kaluza-Klein theory, which can be also regarded as an extension
of the 5d Schwarzschild solution, then we end up with a dramatic
result that the time evolution of the effective (or observed)
universe in the reduced description is in fact a kind of gauge
transformation.

The simplest way of obtaining a 5d cosmological solution (in the
absence of matter) is to make a coordinate transformation
\begin{equation}
t= \int \frac{dR}{[-k(1-{R^2_0}/{R^2})]^{1/2}}
\end{equation}
in the 5d Schwarzschild solution
\begin{equation}
ds^2=[ k(1-\frac{R^2_0}{R^2}) ]^{-1}dR^2 + R^2d\Omega^2_k - [
k(1-\frac{R^2_0}{R^2} )] d\tau^2 ,
\end{equation}
where $R_0$ is some constant, and $d\Omega^2_k$ is the line
element of the 3d volume with constant curvature\footnote{Here we
omit the case $k=0$ for convenience. This, however, dose not ruin
the generality of our discussion.} $k=1,-1$. Upon renaming $\tau
\rightarrow x^5$, we then obtain
\begin{equation}
ds^2=-dt^2+[ R^2_0-k(t-t_0)^2]d\Omega^2_k +
\frac{(t-t_0)^2}{[R^2_0-k(t-t_0)^2]}(dx^5)^2 ,
\end{equation}
the 5d version of the Robertson-Walker metric. Equation(3)
coincides (for $k=1$) with the solution found in ref.\cite{mat}
provided the integral constant $t_0$ is identified with $R_0$.
With this identification the metric(3) describes after obvious
dimensional reduction a closed universe which starts
 to expand from the initial (big-bang) singularity at $t=0$,
reaches maximum radius $R_0$ at $t=R_0$, then collapses to a
point(crunch singularity) as $t\rightarrow 2R_0$. So in this case
both big-bang and crunch singularities are present. For $k=-1$, on
the other hand, the absence of the big-bang singularity is
manifest. In this case, $R_0$ represents the minimum radius
instead of maximum on the contrary to the case $k=1$. The
expansion starts from $R_0$ at $t=t_0$, then continues forever.
Thus, the initial singularity does not exist as long as $R_0$ is a
non-zero constant.

 Though the metric (2) is a vacuum solution in itself, the
effective cosmology immanent in it is not trivial. As discussed in
ref.\cite{mat} metric (3) describes ( in the 4d sector) a
radiation-dominated universe; the dynamics of the fifth dimension
contributes effective radiation source. In fact the analogy
between (3) and the standard Robertson-Walker cosmology becomes
clear once we compare the scale factor $ R^2(t)\equiv R^2_0
-k(t-t_0)^2$ in (3) with the one in standard cosmology which was
found to be \cite{ste}
\begin{equation}
R^2(t)= \left\{\begin{array}{rl} (\sqrt{\kappa
A/3})^2-[t-(\tilde{t_0} + \sqrt{\kappa A/3})]^2 & \mbox{for {\it
k}=1}\\ -(\sqrt{\kappa A/3})^2 + [(t-\tilde{t_0}) + \sqrt{\kappa
A/3}]^2 & \mbox{ for {\it k}=-1} ,
\end{array}\right.
\end{equation}
 where $\kappa \equiv 8 \pi G$, $ \tilde{t_0}$ is an arbitrary
constant, and A is related with energy density of radiation by $
A=\rho R^4 =$constant. From(4) one immediately see that for $k=1$
both $R(t)$'s are identical if we identify $R_0 \leftrightarrow
\sqrt{\kappa A/3}$, and $t_0 \leftrightarrow \tilde{t_0} +
\sqrt{\kappa A/3}$.
 But for $k=-1$ the situation is a little bit different. $R(t)$ in
(4)
starts from the initial singularity at $t=t_0$ instead of starting
from $R_0$. Indeed for $k=-1$ those $R(t)$'s have some different
time dependence near the big-bang as we can see from (3) and (4).
However, it should be noted that both have the same asymptotic
behavior, $R(t) \sim t$ as $t \rightarrow \infty$, which is
typical of radiation-dominated open universe.

Now we get into the main point. First, as an extension of (3) we
introduce an ansatz
\begin{equation}
ds^2= -dt^2 + R^2(t, x^5)d\Omega^2_k + e^{\mu(t,x^5)}(dx^5)^2 .
\end{equation}
Equation(5) is just the (five dimensional) Tolman metric with the
radial coordinate $r$ in the conventional Tolman metric replaced
by the extra coordinate $x^5$. These two metrics (i.e., the metric
in eq.(5) and the conventional Tolman metric) have the same
mathematical form, but their physical contents are quite
different: while the latter describes collapse phase of a dust,
the former a family of evolving universes each of which is
parametrized by $x^5$. Since we are considering an extension of
the vacuum solution the appropriate field equations to be worked
out are the vacuum Einstein equations $R_{AB} =0$, which reduce
after some algebra to a set of independent field equations
\begin{equation}
2\dot{R}^{\prime} - \dot{\mu}R^{\prime} = 0 ,
\end{equation}
\begin{equation}
R\ddot{R} + \dot{R}^{2} + k - e^{-\mu}{R^{\prime}}^2 = 0 ,
\end{equation}
\begin{equation}
(\frac{\dot{R}}{R})^2 + \frac{1}{2}\dot{\mu}\frac{\dot{R}}{R} +
\frac{k}{R^2} - e^{-\mu}[\frac{R^{''}}{R} +
(\frac{R^{\prime}}{R})^2 - \frac{1}{2}\mu^{\prime}
\frac{R^{\prime}}{R} ] = 0 .
\end{equation}
Here, "primes" and "overdots" denote $ \prime \equiv \partial /
\partial x^5$ and $ \cdot \equiv \partial /\partial t$,
respectively, and other equations are simply combinations of these
three equations. The procedure for solving the above field
equations is well known \cite{car}: eq.(6) is integrated to give
\begin{equation}
e^{\mu} = \frac{R^{\prime\/ 2}}{k + f(x^5)} ,
\end{equation}
$f(x^5)$ being an arbitrary function of $x^5$ alone; substituting
(9) in (7) then gives\footnote{In the case of ordinary 4d Tolman
solution the 2nd. term of the r.h.s. of (10) appears to be $\sim
1/R$ ( instead of $ \sim 1/R^2$).}
\begin{equation}
\dot{R}^2 =f(x^5) + \frac{g(x^5)}{R^2}  ,
\end{equation}
$g(x^5)$ being another arbitrary function of $x^5$ alone; finally
by virtue of (9) and (10) the last equation (8) reduces to
\begin{equation}
\frac{1}{R^3}\frac{g^{\prime}}{R^\prime} = 0  ,
\end{equation}
whose obvious solution is
\begin{equation}
g(x^5)=constant \equiv g_0 .
\end{equation}
The procedure then goes differently from here. The most general
solution to eq.(10) is found to be
\begin{equation}
R^2(t, x^5) = R^2_0(x^5) + f(x^5)[t-\hat{t_0}(x^5) ]^2
\end{equation}
with $R_0(x^5)$ defined by
\begin{equation}
 R_0^2(x^5) \equiv -\frac{g(x^5)}{f(x^5)}  ,
 \end{equation}
 and where it is important to note that $\hat{t_0}(x^5)$, which
 has been introduced as an integral constant, is actually not
 simply a constant; it is most generally a function of $x^5$. The
 scale factor (9), on the other hand, is now written as
 \begin{equation}
 e^{\mu(t,x^5)} = \frac{1}{(k+f)}\frac{[R_0R_0^{\prime}
 +(f^{\prime}/2)^2 (t-\hat{t_0})^2
 -f\hat{t_0}^{\prime}(t-\hat{t_0})]^2}{[R^2_0 +f(t-\hat{t_0})^2]}
 ,
 \end{equation}
which, however, diverges as $t\rightarrow \infty$ unless
$f^{\prime}$ is zero. At this point we recall that the size of the
compact extra dimensions in Kaluza-Klein theories is closely
related with observed constants (electric charge or Newton's
constant for instance) of the 4d effective theory \cite{gro}. So
it is customary in Kaluza-Klein cosmology to assume that the
radius of the compact dimension should be constant at least
asymptotically; we therefore require: $e^{\mu(t,x^5)} \rightarrow
constant$ as $ t \rightarrow \infty$. This requirement then
immediately implies
\begin{equation}
f(x^5) = constant \equiv f_0  ,
\end{equation}
\begin{equation}
\hat{t_0}^{\prime}(x^5) = constant \equiv -\alpha ,
\end{equation}
and, by (12) and (16), $R^2_0(x^5)$ in (14) now becomes
\begin{equation}
R^2_0(x^5) = -\frac{g_0}{f_0} =constant \equiv R^2_0  .
\end{equation}
In particular this constant $R_0$ is a turning point in the
classical motion described by (10); note that the kinetic energy
term $\dot{R}^2$ vanishes when $R^2(t,x^5) =-g/f=R^2_0$. Thus the
constant $R_0$ in (18) is identified with $R_0$ in the metric(3).
After all this we find the solution which meets the given
requirement to be written as
\begin{equation}
ds^2=-dt^2 + R^2(t,x^5)d\Omega^2_k + \phi(t,x^5) (dx^5)^2
\end{equation}
with
\begin{equation}
R^2(t,x^5) =R^2_0 +f_0(t-\hat{t_0})^2  ,
\end{equation}
\begin{equation}
 \phi(t,x^5)  \equiv e^{\mu(t,x^5)}
 =\frac{f^2_0(t-\hat{t_0})^2}{R^2_0 + f_0(t-\hat{t_0})^2}  .
 \end{equation}
 Again $R_0$ and $f_0$ in (20) and (21) are constants, but
 $\hat{t_0}$ is a function of $x^5$ :
 \begin{equation}
 \hat{t_0}(x^5) = t_0 -\alpha x^5 \: , \: (t_0 =constant) ,
 \end{equation}
 where the constant $\alpha$ has been taken to be
 \begin{equation}
 \alpha = \pm (k + f_0) ^{1/2}  .
 \end{equation}
Obviously eq.(19) is a generalization of the metric (3); one can
see that (19) reduces to (3) for $f_0 \rightarrow -k$. In fact the
solution(19) is cosmologically the unique extension of (3). The
constant $\alpha$, on the other hand, is real for $f_0\geq -k$,
but it is imaginary for $f_0<-k$ and in this case we obtain 'two
time physics' \cite{bar} by replacing $x^5 \rightarrow i\tau$
($\tau$ being the extra time).

The extended solution (19) has a remarkable feature. It is gauge
equivalent to the static (in the 4d sense) solution. Let us recall
that in 5d Kaluza-Klein theory a local $U(1)$ gauge transformation
takes the form of the special case of the general coordinate
transformation
\begin{eqnarray}
\tilde{x}^{\alpha} & = & x^{\alpha} ,\\
 \tilde{x}^5 & = & x^5 + \kappa \Lambda (x^{\alpha}). \nonumber
 \end{eqnarray}
 Indeed the line element of the 5d Kaluza-Klein
 theory is assumed to take the form
 \begin{eqnarray}
 ds^2_5 & = & {}^{5}{\rm g}_{MN}(x^A)dx^{M}dx^{N} \\
  & = & {}^{4}{\rm g}_{\mu \nu}(x^{\alpha},x^5)dx^{\mu}dx^{\nu} +
  \phi(x^{\alpha},x^5)[dx^5 + \kappa A_{\mu}(x^{\alpha},x^5)dx^
{\mu}]^2 ,\nonumber
  \end{eqnarray}
  and using the standard transformation law for the metric tensor
  ${}^5{\rm g}_{MN}(x^{A})$ one can show that the 4d
  field quantities transform, respectively, as
  \begin{eqnarray}
  {}^4{\rm {\tilde{g}}}_{\mu \nu}(x^{\alpha},\tilde{x}^5)
  &=&{}^4{\rm g}_{\mu \nu} (x^{\alpha}, x^5) ,\nonumber \\
  \tilde{A}_{\mu}(x^{\alpha},\tilde{x}^5)&=&A_{\mu}(x^{\alpha},x^5)
  -\partial_{\mu} \Lambda(x^{\alpha}), \\
  \tilde{\phi}(x^{\alpha},\tilde{x}^5) &=& \phi(x^{\alpha},x^5)
   \nonumber
  \end{eqnarray}
  under (24), which leads us to believe
  that  the transformation (24) is associated with the local $U(1)$ 
gauge
  transformation, and consequently $A_{\mu}(x^{\alpha},x^5)$ can be
  interpreted as a $U(1)$gauge boson after dimensional reduction .
  Also  one can readily see that the line element (25) is
  invariant under (24) and (26), meaning that the transformation
  (24) is really a symmetry of the theory. But it should be 
mentioned
  that the translation along the fifth coordinate is not an isometry
  transformation here because $\phi(x^{\alpha},x^5)$ and other 
fields depend on
  $x^5$; i.e., the space structure is not quite cylindrical, and
  the 'cylindricity' condition fails to hold in this case \cite
{ein}. But
  still, (24) is a symmetry of the theory.
  Now let us apply the above gauge transformation to
  the metric (19). If we take
  \begin{equation}
  \Lambda(x^{\alpha}) =\frac{1}{\alpha \kappa}( t-t_0)  ,
  \end{equation}
  then $t - \hat{t}_0$ becomes simply $\alpha \tilde{x}^5$ , and 
(19)
  reduces to
  \begin{equation}
  d\tilde{s}^2 =-dt^2 + \tilde{R}^2(\tilde{x}^5) d\Omega^2_k +
  \tilde{\phi}(\tilde{x}^5)[d\tilde{x}^5 + \kappa \tilde{A}_0(\tilde
{x}^5)dt]^2
  \end{equation}
  with
  \begin{eqnarray}
  \tilde{R}^2(\tilde{x}^5)&=& R^2_0 + \alpha^2 f_0(\tilde{x}^5)^2
  , \nonumber \\
  \tilde{\phi}(\tilde{x}^5)&=& \frac{\alpha^2 f^2_0 (\tilde{x}^5)^2}
   {R^2_0 + \alpha^2 f_0 (\tilde{x}^5)^2} ,
    \\
\tilde{A}_0(\tilde{x}^5)&=& -\frac{1}{\alpha \kappa}
 = constant.   \nonumber
\end{eqnarray}
This result is remarkable. Being time independent the gauge
transformed metric (28) is associated with a static universe,
while $ds^2$ in (19) describes an evolving universe in the reduced
description, which means that the evolving universe is gauge
equivalent to the static universe with constant gauge potential.
This result is analogous to the Higgs mechanism where the massless
Goldstone boson is eaten up (by the gauge transformation)by photon
which, as a result, acquires a mass. In our case the 'time' (or
time degree of freedom) is eaten up (or absorbed into the fifth
dimension) by the gauge transformation, and as a result the
universe acquires a pure gauge potential. The fact that the
acquired potential is a pure gauge is not difficult to understand
because the metric (19) does not contain any gauge potential term;
i.e., $A_{\mu}(x^{\alpha},x^5) =0$ in (26), so
$\tilde{A}_{\mu}(x^{\alpha},\tilde{x}^5)$ has to be a pure gauge.
In fact the appearance of the pure gauge is more manifest in the
Schwarzschild form of the metric (19)(or equivalently, the metric
(5)). Using (9) and (10) together with (20) and (21) we obtain
from (5):
\begin{equation}
ds^2=(k +f_0\frac{R^2_0}{R^2})^{-1}dR^2 + R^2d\Omega^2_k
-(k+f_0\frac{R^2_0}{R^2})[dx^5 +A_R(R)dR]^2
\end{equation}
with
\begin{equation}
A_R (R)=-\frac{\alpha}{[f_0(1-R^2_0/R^2)]^{1/2}(k+f_0 R^2_0/R^2)}.
\end{equation}
The potential $A_R(R)$ is a pure gauge since
$\overrightarrow{\nabla} \times \overrightarrow{A}$ is obviously
zero. Also note that eq.(30) is not exactly the Schwarzschild
metric, but it goes to (2) as $f_0 \rightarrow -k$, so it is an
extension of the Schwarzschild metric.

To proceed, we now ask how each universe described by (19) and its
gauge transformed version (28) looks to the low-energy physicists
who can not make enough power to probe the fifth dimension. For
those physicists the observed result will be simply the average
over $x^5$ (or $\tilde{x}^5$); i.e. they can only observe $n=0$
massless mode in the expansion
\begin{equation}
g_{MN}(x^{\alpha},x^5) =\sum_{-\infty}^{\infty}
g^{(n)}_{MN}(x^{\alpha})e^{inx^5/R_{c}}  ,
\end{equation}
and similarly for $\tilde{g}_{MN}(x^{\mu},\tilde{x}^5)$. In
eq.(32), $R_{c}$ represents the compactication radius of the fifth
dimension, and $x^5$ (and $\tilde{x}^5$) is periodic with period
$2\pi R_{c}$; i.e., $0 \leq x^5 \leq 2\pi R_{c}$, $x^5 \sim x^5
+2\pi R_{c}$, similarly for $\tilde{x}^5$. The reduced line
elements of (19) and (28), which retain only $n=0$ mode, are
found, respectively, to be
\begin{equation}
ds^2_{(0)}=-dt^2 + [R^{(0)} (t)]^2d\Omega^2_{k} +
\phi^{(0)}(t)(dx^5)^2
\end{equation}
with
\begin{equation}
[R^{(0)}(t)]^2=[R^2_0 +\frac{f_0}{3}(\alpha \pi R_c)^2]
+f_0[(t-t_0) + (\alpha \pi R_c)]^2 ,
\end{equation}
\begin{eqnarray}
\phi^{(0)}(t)& =& f_0 -f_0 \frac{R_0}{2\pi R_c \alpha \sqrt{f_0}}
\{\tan^{-1}[\frac{2\pi R_c \alpha \sqrt{f_0}}{R_0} +
\frac{\sqrt{f_0}}{R_0}(t-t_0)] \nonumber \\
 & &-\tan^{-1}[\frac{\sqrt{f_0}}{R_0}(t-t_0)]\} ,
\end{eqnarray}
and
\begin{equation}
d\tilde{s}^2_{(0)} =-dt^2 +[\tilde{R}^{(0)}]^2d\Omega^2_{k} +
\tilde{\phi}^{(0)}[d\tilde{x}^5 + \kappa \tilde{A}^{(0)}_0dt]^2
\end{equation}
with
\begin{equation}
[\tilde{R}^{(0)}]^2 =[R^{(0)}(t_0)]^2 =R^2_0 +
\frac{4}{3}f_0(\alpha \pi R_c)^2 =constant ,
\end{equation}
\begin{equation}
\tilde{\phi}^{(0)} =\phi^{(0)}(t_0) =f_0-f_0\frac{R_0}{2\pi R_c
\alpha \sqrt{f_0}}\tan^{-1} \frac{2 \pi R_c \alpha
\sqrt{f_0}}{R_0} = constant,
\end{equation}
\begin{equation}
\tilde{A}^{(0)}_0 = -\frac{1}{\alpha \kappa } = constant .
\end{equation}
The universe described by (33) is still a radiation-dominated
universe (in the 4d sector), but a little different\footnote{But
reader can easily check that the line element (33) reduces to (3)
for $f_0 \rightarrow -k$ (or $\alpha \rightarrow 0$).} from that
described by (3). In particular the constant $R^2_0$ in (34)
admits an addition term $f_0(\alpha \pi R_c)^2/3$. This is of
interest particularly in the case $k=-1$. Recall that the constant
$f_0$ must satisfy $f_0\geq -k$ in order for $\alpha$ to be real
in eq.(23). So $f_0$, and consequently the additional term
$f_0(\alpha \pi R_c)^2/3$ should be positive for $k=-1$, which
implies that the big-bang singularity does not exist even in the
case where the free parameter $R_0$ is set equal to zero.

 As mentioned above the line elements (33) and (36)
describe the effective universes the low-energy physicists can
observe, and we see that those universes are entirely distinct
from one another; one is static, another is evolving. From this
one can deduce an important result that two universes mutually
gauge equivalent at the level of full theory retaining all $n\neq
0$ modes can appear to be totally different universes to the
low-energy physicists. In other words, an evolving universe
observed at low energy level may be a different expression of the
static universe with non-zero gauge potential. That is, choosing a
different gauge corresponds to having a different universe with a
different time evolution and a different gauge potential. For the
universe described by (33), the 'time' comes into play in the
gauge $A_0=0$. This is a mystery. Why has a certain observed
universe had to choose the corresponding particular gauge? Or,
what has made it choose that particular gauge? Unfortunately, we
do not know the answer. The only thing we can say is that the
result obtained in this paper is clearly implicated in the
existence of the Kaluza-Klein excitations, and can not be deduced
without considering $x^5$-dependent solutions.

\end{document}